\documentclass[
    aps,prl,twocolumn,
	groupedaddress,superscriptaddress,
	amsfonts,amssymb,amsmath,
	citeautoscript,longbibliography,
	letterpaper, nofootinbib
	]{revtex4-2}

\usepackage{svg}
\usepackage{graphicx}
\usepackage{xcolor}
\usepackage{soul}
\usepackage{bbold}
\usepackage{siunitx}
\usepackage{amsmath,amssymb}
\usepackage{braket}
\usepackage{url}

\usepackage{microtype}
\usepackage[]{newtxtext}
\usepackage[]{newtxmath}
\usepackage{booktabs}
\usepackage{colortbl}
\usepackage{makecell}
\usepackage{silence}
\usepackage[hidelinks]{hyperref} 

\begin{document}

\title{Non-Hermiticity-induced chirality imbalance of Weyl Landau levels}

\author{Sachin Vaidya$^\bigstar$}
\email{svaidya1@mit.edu}
\affiliation{Department of Physics, Massachusetts Institute of Technology, Cambridge, Massachusetts 02139, USA}
\author{Alaa Bayazeed$^\bigstar$}
\email{bayazeed@rptu.de}
\affiliation{Physics Department and Research Center OPTIMAS, RPTU University Kaiserslautern-Landau, Kaiserslautern D-67663, Germany}
\author{André Grossi Fonseca}
\affiliation{Department of Physics, Massachusetts Institute of Technology, Cambridge, Massachusetts 02139, USA}
\author{Adolfo G. Grushin}
\affiliation{Donostia International Physics Center, P. Manuel de Lardizabal 4, 20018 Donostia-San Sebastian, Spain}
\affiliation{IKERBASQUE, Basque Foundation for Science, Maria Diaz de Haro 3, 48013 Bilbao, Spain}
\affiliation{\small Universit\'e Grenoble Alpes, CNRS, Grenoble INP, Institut N\'eel, 38000 Grenoble, France \\
$^\bigstar$ denotes equal contribution}
\author{Marin Solja\v{c}i\'c}
\affiliation{Department of Physics, Massachusetts Institute of Technology, Cambridge, Massachusetts 02139, USA}
\author{Christina J\"{o}rg}
\email{cjoerg@rptu.de}
\affiliation{Physics Department and Research Center OPTIMAS, RPTU University Kaiserslautern-Landau, Kaiserslautern D-67663, Germany}

\date{\today}

\begin{abstract}
Weyl semimetals obey a global chirality constraint: the net chiral topological charge and any associated chiral spectral flow must vanish, as required by the Nielsen–Ninomiya theorem. Under magnetic fields, this constraint manifests through counter-propagating zeroth Landau levels associated with Weyl nodes of opposite chirality. Here, we experimentally demonstrate how non-Hermiticity can reshape this balance in a synthetic photonic Weyl semimetal. Using engineered gauge fields in one-dimensional multilayer structures, we realize both homogeneous and axial magnetic fields and directly probe the resulting Landau-level spectra. While a homogeneous field produces the expected chirality-balanced zeroth Landau levels, an axial field spatially separates the compensating chiral channels: co-propagating bulk pseudo-Landau levels carry one chirality, whereas the opposite chirality resides in boundary-localized surface states. We show that radiative boundary loss selectively suppresses these surface states, removing them from the long-lived observable spectrum and producing an experimentally accessible chirality imbalance. By reducing boundary loss, we recover the hidden chiral channel and reveal its surface-state origin. These results show that non-Hermiticity, present naturally in photonics, can control and relax fundamental chirality constraints in topological systems, enabling access to otherwise forbidden spectral responses.
\end{abstract}

\maketitle
\section{I. Introduction}
Chirality is a fundamental organizing principle in physics, characterizing excitations and responses that are distinguishable from their mirror images. Its relevance extends from microscopic laws to emergent collective phenomena. In particle physics, chirality is central to the weak interaction, which distinguishes left- and right-handed fermions~\cite{peskin2018introduction, tong2018gauge}. In condensed matter and photonic systems, chiral structure appears in magnetic spin textures~\cite{yang2021chiral}, superconducting vortices~\cite{blatter1994}, chiral crystals~\cite{Chang2018}, structured electromagnetic fields~\cite{forbes2019structured}, and topological quasiparticles~\cite{bradlyn2016beyond, yan2017topological}. Among the most prominent examples are Weyl fermions, which carry a chiral topological charge, the Chern number, and arise as low-energy excitations near linear band degeneracies in three-dimensional structures~\cite{Xu2015, Lv2015, Lu2015, Armitage2018, vaidya2020observation, jorg2022observation}. In each of these settings, chirality is not merely a geometric descriptor, but a property tied to symmetry, topology, and dynamical response.

The physical significance of chirality is most clearly seen when it controls measurable observables. In Weyl semimetals, the chirality of a Weyl node governs a range of spectroscopic and transport phenomena, including chiral Fermi-arc surface states~\cite{Wan2011}, optical responses such as the circular photogalvanic effect~\cite{deJuan2017}, and anomaly-related transport under electromagnetic fields~\cite{nielsen1983adler}. A particularly direct manifestation occurs in a magnetic field, where the Weyl spectrum reorganizes into Landau levels. While higher Landau levels are nonchiral, the zeroth Landau level is chiral with its direction of propagation fixed by the chirality of the Weyl node~\cite{nielsen1983adler}. The slope of this branch therefore provides a direct spectral signature of topological handedness.

This direct observability of chirality is accompanied by stringent global constraints. In local Hermitian lattice systems with a periodic momentum space, the Nielsen–Ninomiya theorem~\cite{Nielsen1981.1, Nielsen1981.2} forbids a non-vanishing net chirality of fermions. In Weyl semimetals, this requires Weyl nodes to occur in pairs whose topological charges sum to zero, so that a Weyl node of one chirality is accompanied by a partner of opposite chirality. The same constraint is reflected in the Landau-level spectrum: under an ordinary magnetic field, the zeroth Landau levels associated with opposite Weyl nodes disperse in opposite directions, ensuring that the net chiral spectral flow remains balanced. In an axial field, such as one induced by a strain, both zeroth Landau levels may have the same chirality and in this case, the surface states provide the required channels of opposite chirality to satisfy the aforementioned constraint~\cite{pikulin2016chiral, grushin2016inhomogeneous, behrends2019landau}. Thus, chirality in lattice Weyl systems is a fundamental property that is both locally observable and globally constrained.

The assumptions underlying this balance of chirality, however, are not unchangeable. In particular, Hermiticity can be relaxed in open, driven, or dissipative systems, where coupling to external degrees of freedom gives modes finite lifetimes. Non-Hermitian systems have been shown to modify topological spectra in ways that have no direct analogue in closed Hermitian lattices, including the redistribution, broadening, or selective suppression of spectral branches~\cite{bergholtz2021exceptional, okuma2023non}. Photonic systems provide a natural setting for this physics where non-Hermitian effects become relevant because radiation, absorption, and engineered loss are intrinsic experimental ingredients rather than perturbations to be eliminated. This raises a fundamental question: how is chirality conservation affected in open Weyl systems where chiral modes may be selectively affected by non-Hermiticity?

Here we experimentally demonstrate that non-Hermiticity can produce an observable chirality imbalance in a Weyl system by selectively suppressing the spectral channels that restore chirality conservation. Using a synthetic photonic Weyl semimetal with programmable gauge fields, we first realize a homogeneous magnetic field and observe the conventional chirality-balanced zeroth Landau levels associated with Weyl nodes of opposite topological charge. We then engineer an axial magnetic field that spatially separates the compensating chiral spectral flow into distinct sectors: co-propagating zeroth pseudo-Landau levels in the bulk and counter-propagating topological surface states at the boundaries. While these surface states restore the global chirality balance required by the Nielsen--Ninomiya theorem in a closed Hermitian system, we show that radiative boundary loss in the open photonic structure selectively broadens and suppresses them, removing the compensating chiral channel from the long-lived observable spectrum. The resulting transmission spectrum therefore exhibits an apparent net chirality despite originating from a lattice system with vanishing total topological charge. By reducing the boundary loss, we recover the hidden surface-state contribution and directly reveal its role in restoring chirality balance. Our results establish non-Hermiticity and gauge fields as a powerful route for manipulating chiral spectral flow and accessing topological responses beyond those observable in closed Hermitian systems.

\begin{figure}[]
\centering
\includegraphics[width=\linewidth]{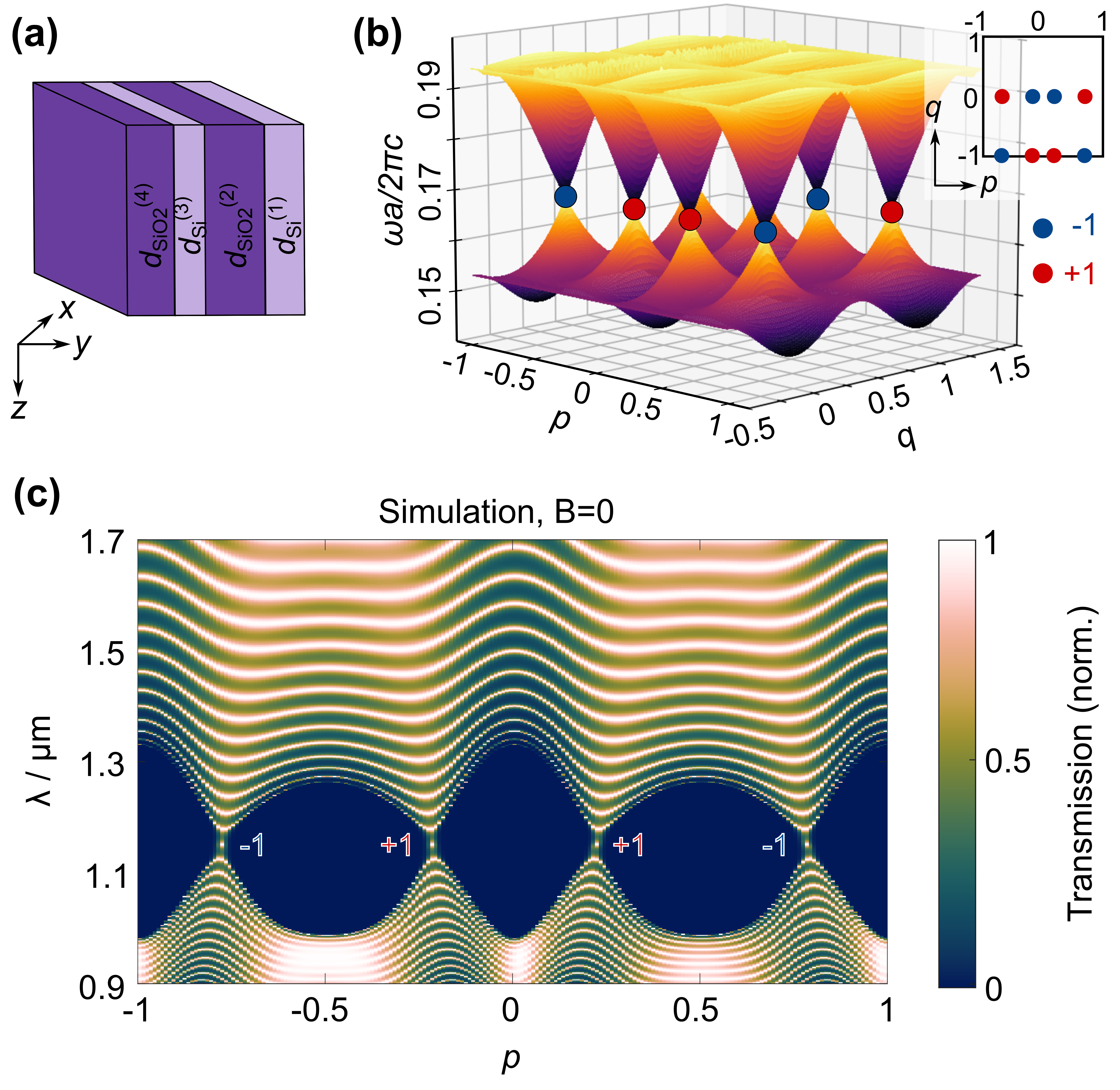}
\caption{ (a) Schematic of one unit cell of the photonic multilayer stack with alternating Si and SiO$_2$ layers. (b) Band structure of frequency ($\omega a/2\pi c$) vs. the two synthetic momenta $p$ and $q$, at $k_y=0$, showing eight Weyl points between bands 2 and 3 with topological charges +1 (red) and -1 (blue). (c) Transfer-matrix simulation for $20$ unit cells of the multilayer stack at $q=0$, $k_y=0$. Along this cut, four Weyl points with charges $-1, +1, +1, -1$ (from left to right) are visible. 
}
\label{fig:1}
\end{figure}

\section{II. Photonic Weyl semimetal in multilayer structures}

We realize a photonic Weyl semimetal using a family of one-dimensional multilayer Bragg stacks composed of alternating silicon (Si) and silicon dioxide (SiO$_2$) layers. Although each individual sample is periodic only along the physical propagation direction, we introduce two additional periodic parameters to modulate the layer thicknesses, which serve as additional synthetic momenta~\cite{Wang2017, nguyen2023fermi, fonseca2024weyl, linn2026topological, vaidya2023reentrant}. This dimensional-extension approach allows a simple multilayer optical structure to emulate the band topology of a three-dimensional Weyl system. 

Light propagation through the multilayer at normal incidence is described by the one-dimensional Maxwell eigenvalue equation~\cite{Joannopoulos2008}
\begin{equation}
    -\partial_y \left[\frac{1}{\epsilon_{p,q}(y)} \partial_y \mathcal{H}(y)\right]
    =
    \left(\frac{\omega}{c}\right)^2 \mathcal{H}(y),
    \label{eq:maxwell}
\end{equation}
where $\epsilon_{p,q}(y)$ is the spatially dependent dielectric permittivity along the stacking direction $y$, $\omega$ is the optical frequency eigenvalue, $c$ is the speed of light in vacuum, and $\mathcal{H}(y)$ is the transverse magnetic field eigenmode. At normal incidence, the transverse electric and transverse magnetic polarizations are degenerate and obey the same scalar equation. The periodic dielectric profile $\epsilon_{p,q}(y)$ is determined by the sequence of layer thicknesses $d^{(n)}$ within the stack that depend on two parameters $p, q$ [Fig.~\ref{fig:1}a]. In particular, we choose the thicknesses of the two silicon and two silica layers within each unit cell to be the following periodic functions:
\begin{align}
    d_{\mathrm{Si}}^{(1)} &= \frac{2}{3}d_0\left[1+\sin(\pi p)\right],
    &
    d_{\mathrm{SiO}_2}^{(2)} &= \frac{1}{3}d_0\left[1+\sin(\pi q)\right],
    \notag \\
    d_{\mathrm{Si}}^{(3)} &= \frac{2}{3}d_0\left[1-\sin(\pi p)\right],
    &
    d_{\mathrm{SiO}_2}^{(4)} &= \frac{1}{3}d_0\left[1-\sin(\pi q)\right],
    \label{eq:layer_thicknesses}
\end{align}
where $a = 2d_0$ is the lattice constant that sets the overall length scale. We note that each value of $(p,q)$ represents a single sample with the layer thicknesses in each unit cell given by Eq.~\ref{eq:layer_thicknesses}. 

Together with the Bloch momentum $k_y$ associated with propagation through the periodic multilayer, the 3-tuple $(p, k_y,q)$ defines an effective three-dimensional toroidal parameter space, mapping to the usual momentum space as $(k_x, k_y, k_z) \leftrightarrow (p, k_y, q)$. The resulting eigenvalue problem for frequency $\omega$ is then equivalent to a 3D band-structure problem in this space. For the choice of layer thicknesses given in Eq.~\ref{eq:layer_thicknesses}, the synthetic band structure contains isolated linear degeneracies between photonic bands, which are Weyl nodes. Near each node, the spectrum is locally conical, and the corresponding Bloch eigenstates carry a quantized Chern number of $\pm 1$ [Fig.~\ref{fig:1}b]. The sign of this integer charge defines the chirality of the Weyl node. Because the system is fully periodic in the parameter space, the total chirality must vanish by the Nielsen-Ninomiya theorem: Weyl nodes occur in oppositely charged pairs, as required for any Hermitian lattice realization of Weyl physics. 

The transmission spectrum through these structures (truncated in the $y$-direction) provides a spectroscopic probe of the photonic system in this parameter space. For example, for a fixed value of $q$, measuring the transmission of light as a function of $p$ gives a projection through this synthetic band structure (with $k_y$ projected out from the truncation), with optical frequency (or wavelength) playing the role of the band energy. The simulated transmission spectrum showing Weyl nodes is shown in Fig.~\ref{fig:1}(c). These Weyl nodes provide the starting point for introducing synthetic magnetic fields. 

\begin{figure}[]
\centering
\includegraphics[width=\linewidth]%
{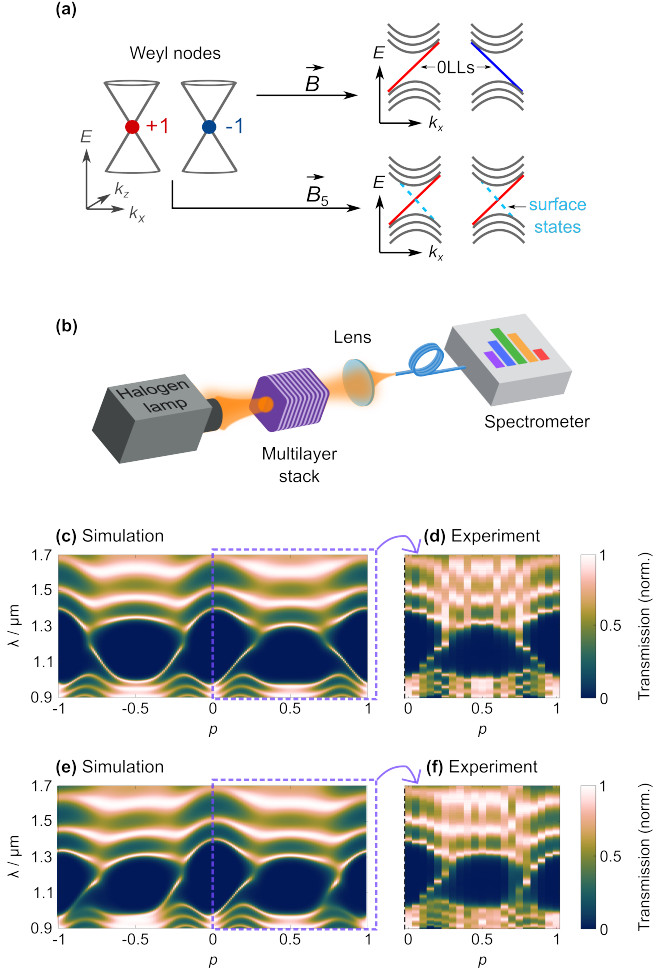}
\caption{ (a) The two oppositely charged Weyl points transform into dispersive zeroth Landau levels (0LLs) under an applied magnetic field. The Landau levels have opposite slopes under a homogeneous $\mathbf{B}$-field and the same slope under an axial magnetic field ($\mathbf{B_5}$). Surface states are indicated by dashed light blue lines. (b) A schematic of the experimental measurement setup: broadband light from a halogen lamp illuminates the multilayer stack and the transmitted spectrum is recorded using a spectrometer. (c) Transfer matrix simulation of the spectrum for $q=0, k_y=0$ under a homogeneous $\mathbf{B}$-field. The zeroth Landau levels inherit their chirality from that of the Weyl nodes shown in Fig.~\ref{fig:2}c. (d) Measurement of the spectrum for positive $p$-values. (e-f) The same for a $\mathbf{B_5}$-field. In this case, all zeroth Landau levels have the same chirality and the spectrum, which is periodic in $p$, exhibits a clear imbalance of chirality.
}
\label{fig:2}
\end{figure}

\section{III. Landau levels under\\a homogeneous magnetic field}

In a conventional Weyl semimetal, a uniform magnetic ($\mathbf{B}$) field quantizes the motion transverse to the field into Landau levels while preserving the momentum parallel to the field. The resulting spectrum contains a set of nonchiral higher Landau levels and a distinguished zeroth Landau level with the sign of its group velocity fixed by the chirality of the corresponding Weyl node. Thus, for a pair of Weyl nodes with opposite topological charge, an ordinary magnetic field produces two zeroth Landau levels with opposite slopes [Fig.~\ref{fig:2}a, top]. These Landau levels have previously been observed in electronic, acoustic and microwave systems~\cite{yuan2018chiral, jia2019observation, peri2019axial, li2025observation, li2025observation_2}.

We next show that a controlled spatial dependence of $q$ along the multilayer stack introduces a homogeneous magnetic field and quantizes the photonic Weyl cones into Landau levels, allowing the chirality of the Weyl nodes to be read out directly from the slopes of the zeroth Landau-level dispersion. We do this by making the synthetic momentum $q$ vary linearly as a function of the discrete unit-cell index $n$ along the propagation direction ($y$). Concretely, for each structure parametrized by $(p,q)$, we replace the constant $q$ by the spatially dependent form given by
\begin{equation}
    q_{\mathrm{hom}}(n)
    = \frac{1}{4}\left(\frac{n-1}{N}-1 \right)
    \qquad
    n \in \{1,2,\ldots,N\},
    \label{eq:q_hom}
\end{equation}
where $n$ is the unit-cell index and there are $N$ total layers in the structure.

This construction can be understood as follows. In our Weyl system, $q$ plays the role of a momentum-like coordinate in the synthetic three-dimensional parameter space $(p,k_y, q)$. The real-space coordinate ($y$) along the multilayer stack is represented by the discrete unit-cell index $n$. Promoting $q$ to a coordinate-dependent quantity through $q_{\mathrm{hom}}(n)$ is thus equivalent to introducing a vector potential ($\mathbf{A}$) that shifts the synthetic momentum coordinate $q$ as a function of position. In particular, the linear dependence of $q_{\mathrm{hom}}(n)$ on $n$ implements the analogue of a Landau-gauge vector potential (at low energies near the Weyl nodes), $A_{\tilde{q}} \propto n$. This is directly analogous to the conventional electronic Landau gauge $A_z = B_0y$ for a homogeneous magnetic field directed along the $x$-direction, $\mathbf{B} = B_0\hat{x}$, with $k_x \parallel \mathbf{B}$ remaining conserved. 

The resulting homogeneous magnetic field is therefore directed along the ``$p$ axis", so that $p$ remains the conserved momentum parallel to the field and the zeroth Landau-level branches disperse along $p$, exactly as the chiral zeroth Landau levels of a Weyl semimetal disperse along the momentum component parallel to an applied magnetic field. As expected, this field couples with the same sign to all Weyl nodes and consequently, the zeroth Landau levels inherit the sign of their group velocities from the respective Weyl node monopole charges. Transfer-matrix simulations of this homogeneous-field configuration in Fig.~\ref{fig:2}c show the Landau level dispersion.

We experimentally probe this spectrum by fabricating a set of multilayer stacks out of Si and SiO$_2$ on glass substrates using the plasma-enhanced chemical vapor deposition (PECVD) technique. We fabricate many structures with different values of $p$ and each having a spatial $q$ dependence given by Eq.~\ref{eq:q_hom}, and measure their transmission spectra using a broadband light source and a spectrometer (see Appendix A for fabrication and measurement details). The measurement setup is schematically shown in Fig.~\ref{fig:2}b. We choose to focus on two Weyl points with opposite charges in the region $p\in[0, 1]$ to illustrate the salient features highlighted by the dashed squares in Fig.~\ref{fig:2}. The measured transmission spectra in Fig.~\ref{fig:2}d reproduce the two oppositely dispersing zeroth Landau-level branches predicted by the transfer-matrix calculation. This chirality-balanced Landau-level spectrum serves as the reference case for the subsequent discussion.

\section{IV. Landau levels under an Axial Magnetic field}

We now turn to the axial-field configuration, where the synthetic magnetic field ($\mathbf{B_5}$) acts with opposite sign on Weyl nodes of opposite chirality. In electronic Weyl systems, such fields can arise from strain-induced shifts of the Weyl-node positions in momentum space or from spatially inhomogeneous magnetization~\cite{ilan2020pseudo}. Unlike an ordinary $\mathbf{B}$ field, which couples identically to both chiralities, an axial field couples with a sign set by the Weyl-node chirality. As a result, the product of chirality and effective field direction entering the zeroth Landau-level velocity can be identical for all Weyl nodes [Fig.~\ref{fig:2}a, bottom]. The corresponding zeroth ``pseudo-Landau levels" therefore disperse in the same direction. In Appendix B we explore the competition between $\mathbf{B}$ and $\mathbf{B_5}$ and the crossover between the homogeneous and axial field regimes for the Landau levels.

In our photonic platform, we implement this axial-field configuration by allowing the synthetic parameter $q$ to depend on both the discrete unit-cell index $n$ and the synthetic momentum $p$. Starting from the homogeneous profile in Eq.~\eqref{eq:q_hom}, we introduce the inhomogeneous modulation
\begin{equation}
    q_{\mathrm{inhom}}(n,p)
    =
    q_{\mathrm{hom}}(n)\cos(\pi p).
    \label{eq:q_inhom}
\end{equation}
Because the effective vector potential now varies and changes sign across the synthetic parameter $p$, Weyl nodes of opposite chiralities experience field configurations with opposite signs. This modulation therefore realizes the photonic analogue of an axial magnetic field ($\mathbf{B_5}$). Transfer-matrix simulations of this structure in Fig.~\ref{fig:2}e show that the zeroth pseudo-Landau-level branches, which counter-propagate in the homogeneous-field case, now acquire the same sign of group velocity. The experimental transmission spectra in Fig.~\ref{fig:2}f reproduce this behavior: over the relevant spectral window, the long-lived bulk modes appear as co-propagating branches in the synthetic momentum coordinate $p$.

\begin{figure}[ht]
\centering
\includegraphics[width=\linewidth]{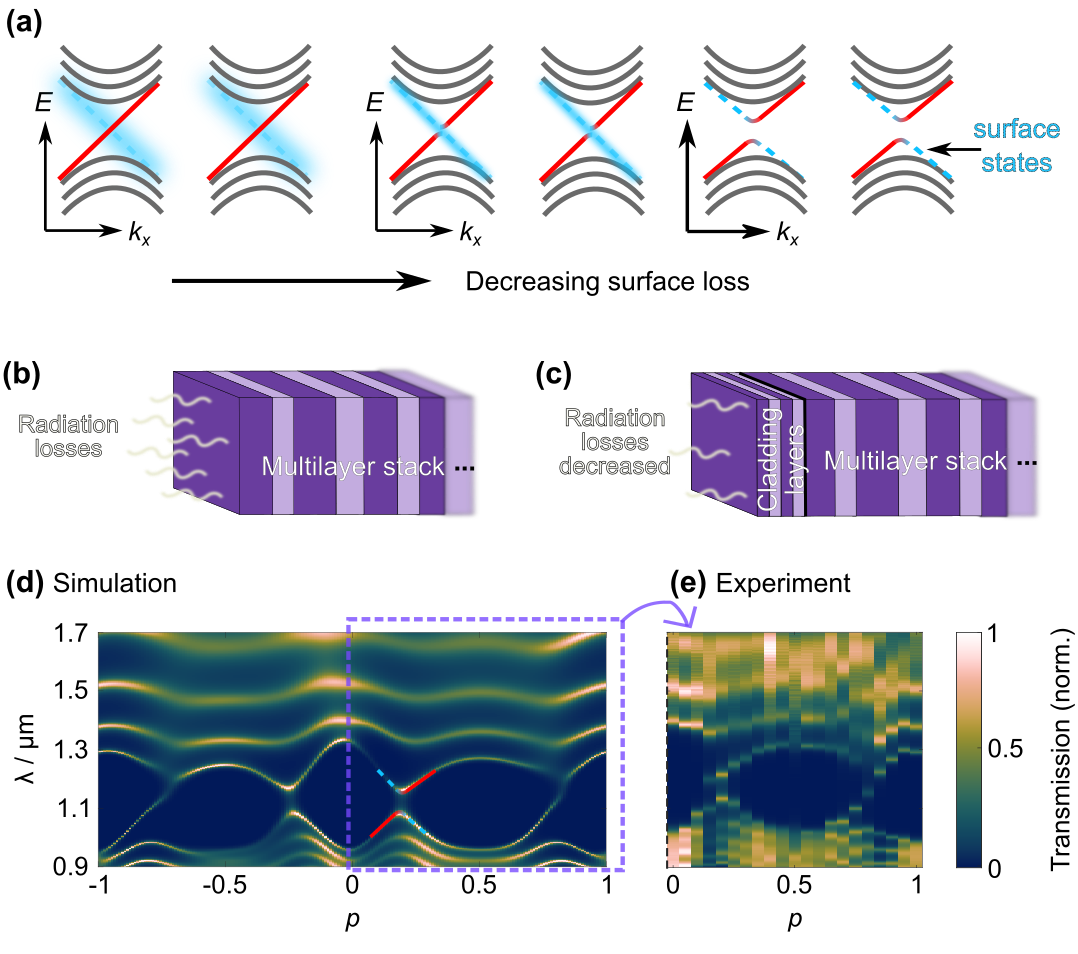}
\caption{(a) Schematic of Landau levels under an axial magnetic field ($\mathbf{B_5}$) and boundary loss: in the transmission spectrum, the surface states (pale blue) are highly broadened and unobservable. Upon decreasing surface loss, they become long-lived and are visible as sharp resonances that hybridize with the zeroth Landau levels (red) in both transfer-matrix simulations (d) and the experiment (e). To reduce the loss, two bilayers of cladding have been added to the top of the multilayer stack, as shown in (b) and (c).}
\label{fig:3}
\end{figure}

At first sight, these co-propagating branches in Fig.~\ref{fig:2}e,f appear to produce an imbalance of chiral spectral flow. In a closed Hermitian lattice system, however, such an imbalance cannot occur in isolation. The resolution is that the chirality is not carried solely by the bulk pseudo-Landau levels. In a finite Weyl system, the co-propagating bulk branches are accompanied by surface states, whose dispersion supplies the compensating spectral flow (see pale blue dashed lines in Fig.~\ref{fig:2}a, bottom). The bulk and surface sectors therefore together satisfy the global chirality constraint~\cite{pikulin2016chiral, grushin2016inhomogeneous, behrends2019landau}. An interesting observation here is that the axial field separates the chiral channels spatially, with one contribution carried by the bulk Landau levels and the opposite contribution carried by surface states localized near the system boundaries.

The multilayer structure, however, is not a closed Hermitian system. It is an open optical system whose boundaries are coupled to external radiation channels in free space. Light can enter and leave through the interfaces with free space and the substrate, and modes localized near these interfaces therefore acquire finite lifetimes. Since the compensating chiral modes in the axial field configuration are surface-localized, they are affected much more strongly by this boundary loss than the extended bulk Landau levels. Due to this loss, the surface states acquire large imaginary eigenvalue components, broaden spectrally, and lose their sharp signatures in the transmission spectrum [Fig.~\ref{fig:3}(a)]. The bulk Landau-levels, by contrast, remain comparatively long-lived and continue to appear as well-defined resonances.

These non-Hermitian boundaries explain the chirality imbalance observed in the transmission spectra in Fig.~\ref{fig:2}e,f. The counter-propagating chiral partners are rendered experimentally invisible by their short lifetimes. The measured spectrum is therefore not the complete Hermitian spectrum, but the long-lived sector selected by the open photonic boundary conditions. In this sector, the observable zeroth pseudo-Landau levels co-propagate, giving rise to a chiral response. This mechanism bears some similarity to the extensively studied Hatano--Nelson model which hosts a complex dispersion relation in which modes with opposite group velocities acquire different amplification or attenuation rates, leading to directional transport and non-reciprocal long-time dynamics~\cite{hatano1996localization}. In a similar way, non-Hermiticity controls which part of the spectral flow remains visible on the relevant time (and frequency linewidth) scales of our experiment. To study this explicitly in a tight-binding setting, we model the system using a lattice Weyl Hamiltonian in a slab geometry with non-Hermitian boundary terms in Appendix C. 

A direct experimental test of the above interpretation is to modify the boundary loss and determine whether the hidden surface branch can be restored in our system. To this end, we add several cladding layers that act as a dielectric Bragg reflector to one side of the multilayer stack [Fig.~\ref{fig:3}c]. The reflector is designed to provide a photonic bandgap near the spectral region of the Weyl points, where the surface contribution is expected, using alternating Si and SiO$_2$ layers satisfying the quarter-wavelength condition. Because the reflector is finite, it does not fully close the system; rather, it partially suppresses radiative leakage at one interface while still allowing optical access to the structure. In the effective non-Hermitian description, this reduces the decay rate of the surface-localized mode at the modified boundary.

Transfer-matrix simulations including the Bragg reflector in Fig.~\ref{fig:3}d show the re-emergence of a surface state of opposite chirality. As its lifetime increases, this formerly broadened surface state becomes sufficiently well defined to hybridize with the bulk spectrum (see Appendix B for the tight-binding model). This hybridization produces an avoided crossing in the transmission map. The same avoided crossing between the surface state and bulk Landau level is observed experimentally in the structures with the Bragg reflector in Fig.~\ref{fig:3}e, while it is absent in the structures without it in Fig.~\ref{fig:2}f. The appearance of this feature only after reducing boundary leakage provides evidence that the missing counter-propagating branch in the axial field case was indeed carried by a lossy surface mode. Importantly, we note that in Fig.~\ref{fig:3}d,e two of the zeroth Landau levels remain unpaired as their surface state partners live on the other surface to which we have not added a cladding. Those surface states therefore remain too broadened to be observable in the spectrum. Adding the cladding to both surfaces would restore the full balance of chirality in this case. 

The axial-field experiment establishes the central contribution of this work. A pseudo-magnetic field creates co-propagating bulk zeroth Landau levels by separating the chiral spectral flow into bulk and boundary sectors. In a Hermitian finite system, the boundary sector supplies the required compensating chirality. In the open photonic multilayer, radiative boundary loss strongly broadens these surface states and removes them from the long-lived transmission spectrum. By partially restoring confinement with a Bragg reflector, the hidden boundary contribution can be recovered. This demonstrates how non-Hermitian boundaries can reshape the manifestation of chirality conservation.

\section{V. Discussion and conclusion}
In conclusion, we have experimentally realized a synthetic photonic Weyl platform in which spatially engineered gauge fields quantize Weyl cones into Landau levels and enable direct control over the chirality and propagation of the associated zeroth modes. By introducing controlled spatial variations of the synthetic parameters, we implemented both homogeneous and momentum-dependent axial gauge fields, directly visualizing the resulting Landau-level structure. We further demonstrated that axial gauge fields and non-Hermitian boundaries from radiative loss provide a mechanism through which modes of one chirality can be preferentially attenuated while the modes of opposite chirality remain within the bulk. This constitutes an effective circumvention of the Nielsen–Ninomiya theorem in the long-lived observable sector, enabled by non-Hermitian boundary loss.

More broadly, the results presented here establish a route towards realizing and directly probing complex multidimensional topological phenomena in an experimentally accessible photonic platform. The present framework offers several conceptual and practical advantages compared to other photonic or condensed-matter realizations. First, the synthetic parameters are directly programmable rather than emergent crystal momenta, enabling gauge fields to be engineered through deterministic spatial control of the multilayer structure itself. This provides direct access to regimes that are difficult to realize electronically, including independently tunable homogeneous and momentum-dependent gauge fields~\cite{hirsbrunner_anomalous, vaidya_quantized}, as well as spatially varying synthetic fields that would require highly nontrivial strain profiles to be incorporated into the structure~\cite{Guglielmon2021, Barsukova2024, Barczyk2024, zhang_airy}. Second, because the system is optical, non-Hermitian phenomena arise naturally through radiative coupling to free space or through deliberately patterned loss and gain. 

The coexistence of programmable dimensions, gauge fields, topology, and controllable non-Hermiticity therefore creates a highly versatile setting for several future directions. The present approach can be generalized to study higher-dimensional topological phases, non-Abelian synthetic gauge fields, topological transport in aperiodic systems, and the manifestations of chirality imbalance in these systems. In addition, the interplay between axial gauge fields and non-Hermiticity suggests new routes towards directional amplification~\cite{wanjura2020topological}, optical isolation~\cite{wang2025topological}, and topological state engineering in open photonic systems~\cite{peano2016topological, cerjan2019experimental}. 

\section{Acknowledgments}
We thank Mikael C. Rechtsman and Georg von Freymann for helpful discussions. We also thank Ellen Bold and Egbert Oesterschulze for the use of their PECVD tool and for their fabrication expertise. C.J. and A.B. gratefully acknowledge financial support from the DFG through SFB TR 185, Project No. 277625399, and the Fulbright-Cottrell Award 2024-25. S.V., A.G.F. and M.S. acknowledge support from the U.S. Office of Naval Research (ONR) Multidisciplinary University Research Initiative (MURI) under Grant No. N00014-20-1-2325 on Robust Photonic Materials with Higher-Order Topological Protection. This material is based upon work also supported in part by the U. S. Army Research Office through the Institute for Soldier Nanotechnologies at MIT, under Collaborative Agreement Number W911NF-23-2-0121. A.~G.~G.~ acknowledges support from the European Research Council (ERC) Consolidator grant under grant agreement No.~101042707 (TOPOMORPH).

\section{Appendix A: Methods}
Simulations of the band structure and calculations of the Chern numbers associated with the Weyl points were performed using the plane-wave expansion (PWE) method, as implemented in the MIT Photonic Bands (MPB) package~\cite{MPB}. Simulations of the transmission spectra were performed using the rigorous coupled-wave analysis (RCWA) method, as implemented in the Stanford S$^4$ package~\cite{StanfordS4}. Since the structure is a one-dimensional multilayer stack, this effectively reduces to a transfer-matrix problem.

All multilayer films were deposited on glass substrates that were ultrasonically cleaned for 15 min successively in acetone, isopropanol, and deionized water. Depositions were carried out using plasma-enhanced chemical vapor deposition (PECVD) in a Plasmalab 80 Plus system (Oxford Instruments). The silicon (Si) layers were deposited from silane (SiH\textsubscript{4}) precursor gas, with ultra-high-purity argon (N60) used as an inert carrier gas under plasma conditions at a substrate temperature of \SI{300}{\degreeCelsius}. Silicon dioxide (SiO\textsubscript{2}) layers were deposited through plasma-enhanced oxidation of SiH\textsubscript{4} using nitrous oxide (N\textsubscript{2}O), while maintaining the same deposition conditions. The target layer thicknesses were converted into deposition times using experimentally calibrated PECVD deposition rates of \SI{48.66}{nm/min} for Si and \SI{72.1}{nm/min} for SiO\textsubscript{2}. Each fabricated sample for the homogeneous field case corresponds to a specific value of the synthetic parameters $(p, q)$. For the axial field case, each fabricated sample corresponds to a specific value of $p$ and an associated $q$-dependent spatial profile encoded directly into the deposition sequence.

\begin{figure}[ht]
\centering
\includegraphics[width=\linewidth]{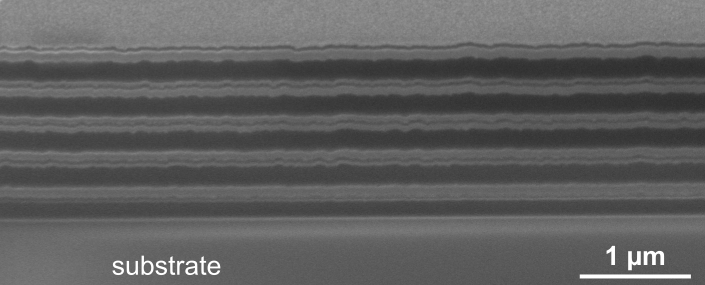}
\caption{Focused ion beam (FIB) cut through a typical multilayer stack revealing the different layers. The silicon layers appear dark gray while the silicon dioxide layers appear light gray.}
\label{fig:appendix_A}
\end{figure}

The deposited layer thicknesses were characterized from focused ion beam (FIB) cross-sectional images (see Fig.~\ref{fig:appendix_A} for a typical example). The total measured multilayer thickness differed from the theoretical design value by approximately \(5\%\), and this experimentally measured thickness was subsequently incorporated into the numerical simulations through an effective calibrated base thickness. The observed deviation likely arises from a combination of fabrication and characterization uncertainties. On the fabrication side, gradual variations in deposition rates during the PECVD process, including precursor depletion, plasma non-uniformity, and temperature fluctuations, can lead to systematic thickness deviations. On the characterization side, uncertainties in the FIB image analysis, such as finite pixel resolution, imperfect interface contrast, and manual determination of layer boundaries, can accumulate across the multilayer stack. The resulting discrepancy therefore remains within the expected uncertainty range of the fabrication and measurement processes.

To experimentally probe photonic analogs of Fermi arc surface states, it is necessary to ensure sufficient confinement of the corresponding surface-localized optical modes. To this end, a dielectric Bragg reflector was deposited on top of the multilayer structure. The reflector was designed to exhibit high reflectivity near \(\lambda=\SI{1200}{nm}\), corresponding to the spectral region where the surface state is predicted to occur. The thicknesses of the alternating Si and SiO\textsubscript{2} layers were determined using the quarter-wave condition with refractive indices \(n_{\mathrm{Si}}=3.575\) and \(n_{\mathrm{SiO_2}}=1.45\). The Bragg reflector consisted of two Si/SiO\textsubscript{2} bilayers (four total layers). Because only two bilayers were employed, the reflector remains partially transmissive rather than forming an ideal high-reflectivity mirror. This controlled leakage enables partial coupling of the surface state to free space, allowing its spectral signature to be observed in transmission measurements while still maintaining sufficient confinement to support the mode. In this regime, the Bragg reflector functions as a leaky top mirror that selectively reveals the surface-localized state.

Optical transmission measurements were performed to characterize the spectral response of the fabricated structures. The samples were illuminated using a broadband halogen light source, and the transmitted signal was collected with a fiber-coupled near-infrared spectrometer (Ocean Insight NIRQuest) operating over the wavelength range from 900 to 1700\,nm. For each sample corresponding to a specific value of \(p\), the transmission spectrum was recorded as a function of wavelength. To maximize collection efficiency, the transmitted beam was focused into the input optical fiber using a collection lens.

\section{Appendix B: Competition Between B and B$_5$}
The homogeneous and axial magnetic fields discussed in the main text represent two limiting cases of a more general situation in which both fields are present simultaneously. For a Weyl node of chirality $\chi=\pm1$, the low-energy theory depends on the effective magnetic field
\begin{equation}
    \mathbf{B}_{\chi}=B+\chi B_{5},
\end{equation}
where $B$ is the ordinary magnetic field and $B_{5}$ is the axial magnetic field. The Landau quantization near each Weyl node is therefore determined by a different effective field strength.

When $|B|\gg|B_{5}|$, the Weyl nodes experience magnetic fields of the same sign and the resulting zeroth Landau levels propagate in directions given by the chiralities of the Weyl nodes. In contrast, when $|B_{5}|\gg|B|$, the effective magnetic fields at the Weyl nodes have opposite signs for opposite chiralities. Because the propagation direction of the zeroth Landau level depends on both the Weyl-node chirality and the sign of the effective field, the two zeroth Landau levels now co-propagate. As discussed in the main text, the compensating counter-propagating channels required by the Nielsen-Ninomiya theorem are supplied by states localized at the system boundaries.

The interpolation between these two regimes is controlled by the condition $B_{\chi}=0$ for a particular Weyl node. At this crossover point the corresponding node no longer experiences Landau quantization. The Landau-level spacing collapses continuously and the spectrum reforms into the standard Weyl cone. Upon crossing this point, the effective field changes sign and the propagation direction of the corresponding zeroth Landau level reverses.

\begin{figure}[ht]
\centering
\includegraphics[width=\linewidth]%
{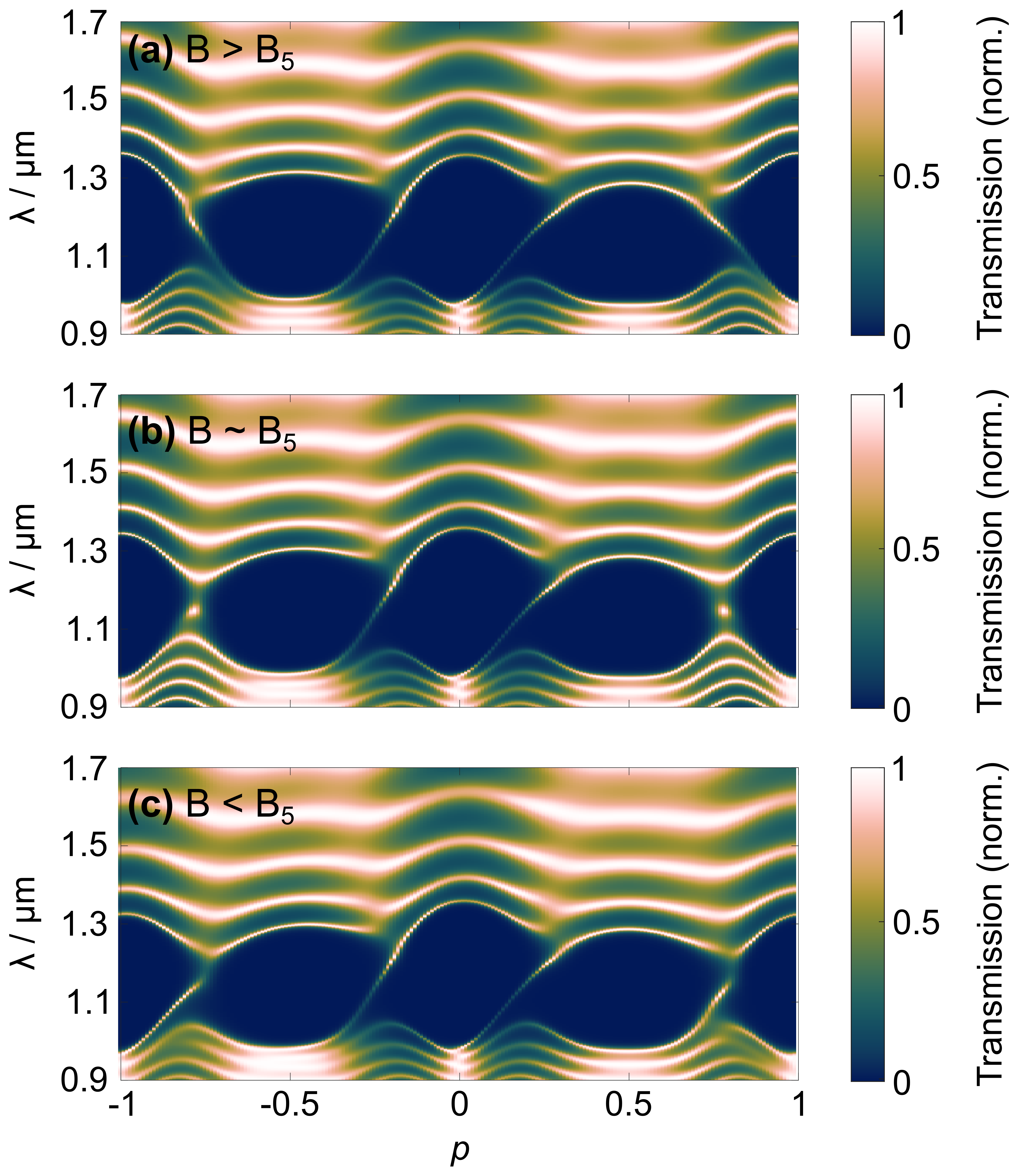}
\caption{Competition between the $B$ and $B_5$-fields. (a) For $B>B_5$ ($\delta=0.3$) the four zeroth Landau levels disperse in opposite directions, whereas for (c) $B<B_5$ ($\delta=0.8$) they disperse in the same direction. (b) At the crossover point $B\approx B_5$ ($\delta=0.6$), the outer two zeroth Landau levels collapse to Weyl points.
}
\label{fig:Appendix_B}
\end{figure}

To explore this in our photonic system, we implement a modulation in the parameter $q$ that consists of both the homogeneous and axial magnetic field terms from Eq.~\ref{eq:q_hom} and Eq.~\ref{eq:q_inhom} of the main text with an interpolation parameter $\delta$. Explicitly, we choose
\begin{equation}
    q_{\mathrm{\delta}}(n)= \delta \cdot q_{\mathrm{hom}}(n)\cos(\pi p) + (1-\delta)\cdot q_{\mathrm{hom}}(n).
\end{equation}
The simulated transmission spectrum for this modulation is shown in Fig.~\ref{fig:Appendix_B} for various values of $\delta$. Here, we observe a crossover between the two regimes at $\delta \sim 0.6$ where two set of Landau levels collapse and the Weyl cones reemerge.

\section{Appendix C: Tight Binding Model}\label{sec:AppendixB}
Here, we examine the spectral broadening of the surface states due to boundary loss in a tight-binding setting. We use a minimal 3D lattice model, with Bloch Hamiltonian
\begin{equation}
\begin{split}
h_{\mathbf{k}}={}&
v[\sin(k_y a)\sigma_x-\sin(k_x a)\sigma_y]\tau_z
+v\sin(k_z a)\tau_y  \\
&+t\sum_i[1-\cos(k_i a)]\tau_x
+va\,\mathbf{u}\cdot\mathbf{b},
\end{split}
\end{equation}
with \(\mathbf{u}=(-\sigma_x\tau_x,-\sigma_y\tau_x,\sigma_z)\), \(v\) the Fermi velocity of the Weyl cones, lattice constant \(a\), and Pauli matrices \(\sigma_\mu\) and \(\tau_\mu\)~\cite{behrends2019anomalous, behrends2019landau, vazifeh2013electromagnetic}. The vector \(\mathbf{b}\) controls the separation of the two Weyl nodes and, at low energies, plays the role of an axial vector potential. We consider open boundary conditions in the \(y\) direction and maintain translation symmetry in the remaining directions. We take the Weyl node separation to be spatially dependent, as \(\mathbf{b}=(b_z,0,B_5y)\), which generates an axial magnetic field \(\mathbf{B}_5=B_5\hat{\mathbf x}\), parallel to the constant node separation in the $x$-direction.

In the presence of this axial magnetic field, the field flips sign for Weyl nodes of opposite chiralities. The bulk zeroth pseudo-Landau levels therefore propagate in the same direction, while the connecting surface states twist in momentum space and provide the counterpropagating branch.

\begin{figure}[]
\centering
\includegraphics[width=1.\linewidth]{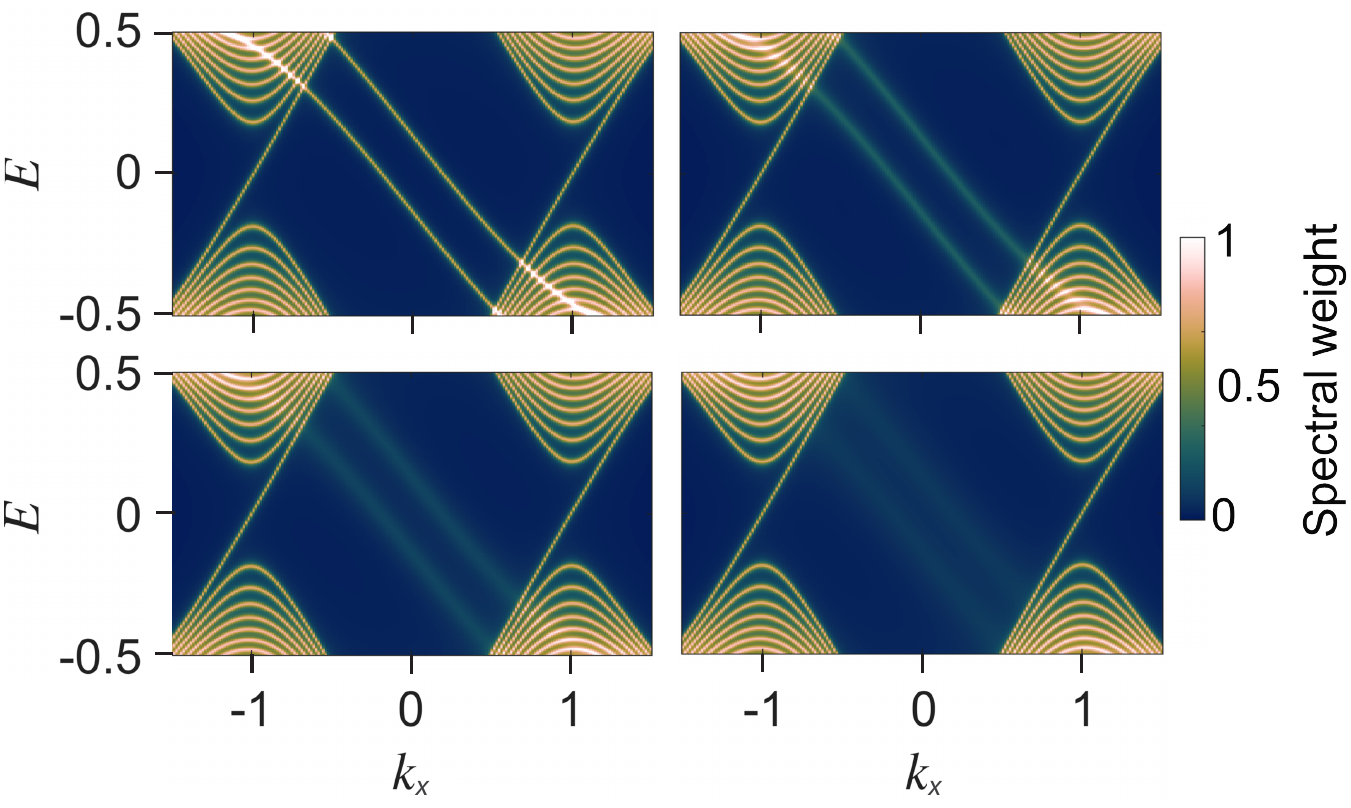}
\caption{
Spectral weight of dispersion along the cut \(k_z=0.19/a\) with $b_z = 1$ of the Weyl semimetal tight-binding model in Eq. (7), under an axial magnetic fields of strength $B_5 = -0.02$ with loss added to surfaces.
The lattice has open boundaries along \(y\), with \(L/a=100\).
The loss increases from left to right, top to bottom, with strength $\eta = 0, 0.02, 0.05, 0.1$, respectively.
}
\label{fig:appendix_C}
\end{figure}

In order to model the radiative surface losses of the 1D photonic crystal, we add to the Hamiltonian a loss term to the surfaces only, which is proportional to the identity in the Pauli basis and with strength $i\eta$.
In the absence of loss ($\eta = 0$), the zeroth pseudo-Landau levels and Fermi arcs counterpropagate as described above, in accordance with the Nielsen--Ninomiya theorem.
However, as the loss increases, the surface Fermi arcs get strongly broadened in their spectral weight until they completely vanish, while the bulk pseudo-Landau levels are largely unaffected, precisely mimicking the chirality imbalance seen in our photonic experiments (Fig.~\ref{fig:appendix_C}).

\bibliography{bibliography}

\end{document}